\documentclass[a4paper]{jpconf}

\usepackage{upgreek}
\usepackage{enumerate}
\usepackage{float}
\restylefloat{table}
\usepackage{graphicx}
\usepackage[all]{xy}
\usepackage{amsmath}
\usepackage{amsthm}
\usepackage{amssymb}
\usepackage[usenames,dvipsnames]{xcolor}

\def\dd{\mathrm{d}}

\newcommand{\Hom}{{\rm Hom}}
\newcommand{\uHom}{\underline{\Hom}}
\newcommand{\ben}{\be}
\newcommand{\een}{\ee}

\newcommand{\eqdef}{\stackrel{{\rm def.}}{=}}
\newcommand{\cA}{{\cal A}}

\newcommand{\cM}{{\cal M}}

\newcommand{\Z}{\mathbb{Z}}

\def\dd{\mathrm{d}}

\def\Pic{\mathrm{Pic}}

\DeclareFontFamily{U}{rsf}{}
\DeclareFontShape{U}{rsf}{m}{n}{<5> <6> rsfs5 <7> <8> <9> rsfs7 <10-> rsfs10}{}
\DeclareMathAlphabet\Scr{U}{rsf}{m}{n}

\def\C{{\mathbb{C}}}

\def\Z{{\mathbb{Z}}}

\def\rk{{\rm{rk}}}

\def\dd{\mathrm{d}}

\def\cO{\mathcal{O}}

\def\Ob{\mathrm{Ob}}
\def\O{\mathrm{O}}

\newcommand{\be}{\begin{equation*}}
\newcommand{\ee}{\end{equation*}}
\newcommand{\beqa}{\begin{eqnarray*}}
\newcommand{\eeqa}{\end{eqnarray*}}

\def\cH{\mathcal{H}}
\def\cT{\mathcal{T}}

\def\MF{\mathrm{MF}}
\def\HMF{\mathrm{HMF}}

\newtheorem{Theorem}{Theorem}[section]
\newtheorem{Lemma}[Theorem]{Lemma}
\newtheorem{Proposition}[Theorem]{Proposition}

\newtheorem{Definition}[Theorem]{Definition}
\theoremstyle{definition}

\def\mod{\mathrm{mod}}
\def\umod{\underline{\mod}}

\def\0{{\hat{0}}}
\def\1{{\hat{1}}}

\newcommand \pd {{\partial}}

\def\rH{\mathrm{H}}
\def\cO{\mathcal{O}}

\def\cN{\mathcal{N}}

\def\rM{\mathrm{M}}
\def\hW{{\hat W}}

\def\hSigma{{\hat \Sigma}}
\def\bk{\mathbf{k}}
\def\ord{\mathrm{ord}}
\def\cZ{\mathcal{Z}}
\def\ZMF{\mathrm{ZMF}}
\def\zmf{\mathrm{zmf}}
\def\hmf{\mathrm{hmf}}
\def\rD{\mathrm{D}}
\def\Sing{\mathrm{Sing}}
\def\rOmega{\mathrm{\Omega}}

\begin{document}

\title{B-type Landau-Ginzburg models with one-dimensional target}

\author{Calin Iuliu Lazaroiu$^1$ and Mehdi Tavakol$^2$}

\address{$^1$ Center for Geometry and Physics, Institute for Basic Science, Pohang 37673,  Korea}
\address{$^2$ School of mathematics and statistics, University of Melbourne, VIC 3010, Australia}

\ead{calin@ibs.re.kr, mehdi.tavakol@unimelb.edu.au}

\begin{abstract} We summarize the main results of our investigation of
B-type topological Landau-Ginzburg models whose target is an arbitrary
open Riemann surface. Such a Riemann surface need not be affine
algebraic and in particular it may have infinite genus or an infinite
number of Freudenthal ends. Under mild conditions on the
Landau-Ginzburg superpotential, we give a complete description of the
triangulated structure of the category of topological D-branes in such
models as well as counting formulas for the number of topological
D-branes considered up to relevant equivalence relations.
\end{abstract}

\section{Introduction}

\

\noindent It is well-known \cite{LG1,LG2} that classical B-type
topological Landau-Ginzburg (LG) models with D-branes can be associated to
any pair $(X,W)$, where $X$ is a non-compact K\"ahler manifold and
$W:X\rightarrow \C$ is a non-constant holomorphic function defined on
$X$. When the canonical line bundle of $X$ is holomorphically trivial
and the critical set of $W$ is compact, it is expected that the
quantization of such models produces a non-anomalous two-dimensional
topological field theory (2d TFT) which obeys the general axioms
introduced in \cite{top} and hence is characterized entirely by an
open-closed {\em TFT datum}, an algebraic structure subject to certain
axioms which encode the sewing constraints of the TFT. As shown in
\cite{LG2}, path integral arguments lead to a proposal for the
open-closed TFT datum of the quantum B-type LG model defined by such a
pair $(X,W)$, a proposal which was clarified and analyzed
mathematically in \cite{lg1,tserre} (see \cite{lgproc} for a brief
summary). When $X$ is Stein and Calabi-Yau, this proposal simplifies
as explained in \cite{lg2} and summarized in \cite{mirela}.

As already pointed out in \cite{lg1}, an interesting particular case
arises when $X$ is complex one-dimensional, i.e. an open Riemann
surface (a non-compact Riemann surface without boundary, which we
assume to be connected).  Since any open Riemann surface is Stein and
holomorphically parallelizable (hence Calabi-Yau), this fits into the
class of models considered in \cite{lg2}, whose TFT datum admits the
simplified description discussed in loc. cit. and summarized in
\cite{mirela}.

In this contribution, we outline our investigation of B-type LG models
with open Riemann surface target \cite{surface}, which in turn relies
on algebraic results derived for a large class of rings in references
\cite{bezout,edd}. We make free use of certain concepts and
terminology which were introduced in \cite{top} and
\cite{LG1,LG2,lg1,lg2} and which are summarized in
\cite{lgproc} and \cite{mirela}.

The open Riemann surfaces considered in this contribution are fully
general. In particular, they can have infinite genus and an infinite
number of Freudenthal ends\footnote{An open Riemann surface coincides
with the analyticization of a complex affine curve iff it has finite
genus and a finite number of ends, in which case it can be obtained by
removing a finite set of points from a compact Riemann surface.}. As
such, the topological LG models discussed herein are the most general
non-anomalous B-type LG models with non-singular complex
one-dimensional target.

\section{Open Riemann surfaces}

\

\noindent In this section, we recall some well-known facts regarding
open Riemann surfaces.

\begin{Definition}
A {\em Riemann surface} (in the sense of Weyl-Rad\'o) is a Hausdorff,
borderless and connected complex manifold $\Sigma$ such that
$\dim_\C\Sigma=1$.
\end{Definition}

\begin{Theorem}[Rad\'o]
The following mutually-equivalent statements hold for any Riemann surface $\Sigma$:
\begin{enumerate}[(a)]
\item $\Sigma$ is paracompact.
\item $\Sigma$ has a countable basis.
\item $\Sigma$ is countable at infinity.
\end{enumerate}
\end{Theorem}

\begin{Definition}
A Riemann surface is called {\em open} if it is non-compact. 
\end{Definition}

\subsection{The topological type of open Riemann surfaces}

\

\noindent Unlike the compact case, the topological classification of
open Riemann surfaces is quite involved, since such a surface can have
infinite genus as well as an infinity of Freudenthal ends
(a.k.a. ``ideal points''). The ends of Riemann surfaces were studied
in the classical work of Ker\'ekj\'art\'o \cite{Kerekjarto} and
Stoilow \cite{Stoilow} (see \cite{Richards} for a summary in modern
mathematical language).

\begin{Definition}

\

\begin{itemize}
\item The {\em transfinite genus} of $\Sigma$ is the cardinal number: 
\be
g(\Sigma)=\frac{1}{2}\rk_\Z \rH^1(\Sigma,\Z)~~.
\ee
\item The {\em ideal boundary} $\pd_\infty\Sigma$ of $\Sigma$ is the
  set of Freudenthal ends of $\Sigma$, endowed with its natural
  topology.
\end{itemize}
\end{Definition}

\begin{Proposition}[Ker\'ekj\'art\'o]
The following statements hold for any open Riemann surface\,$\Sigma$:
\begin{enumerate}[1.]
\item $g(\Sigma)$ is finite or countable.
\item $\pd_\infty\Sigma$ is a finite or countable compact Hausdorff
  topological space which is totally disconnected.
\end{enumerate}
\end{Proposition}

\paragraph{\bf Remarks.}

\

\begin{enumerate}[1.]
\item $\pd_\infty\Sigma$ can be a Cantor space.
\item Adjoining $\pd_\infty\Sigma$ to $\Sigma$ produces the so-called {\em
  Ker\'ekj\'art\'o-Stoilow compactification} $\hSigma$ of $\Sigma$ (see \cite{Kerekjarto, Stoilow, Richards}). 
\end{enumerate}

\noindent There exists a natural disjoint union decomposition
$\pd_\infty\Sigma=\pd_\infty^1\Sigma\sqcup \pd_\infty^2\Sigma$ where
the ends belonging to $\pd_\infty^1\Sigma$ and $\pd_\infty^2\Sigma$
are called of {\em first} and {\em second} kind.

\subsection{The complex geometry of open Riemann surfaces}
\label{subsec:RScx}

\

\noindent The following statements summarize classical results due to Ker\'ekj\'art\'o and Stoilow:

\begin{Theorem}[Ker\'ekj\'art\'o] Let $\Sigma$ and $\Sigma'$ be two
open Riemann surfaces. Then the following statements are equivalent:
\begin{enumerate}[(a)]
\itemsep 0.0em
\item $\Sigma$ and $\Sigma'$ are unoriented-homeomorphic.
\item $g(\Sigma)\!=\!g(\Sigma')$ and there exists a homeomorphism
$\psi\!:\!\pd_\infty\Sigma\!\rightarrow\! \pd_\infty\Sigma'$ such that
$\psi(\pd_\infty^1\Sigma)\!=\!\pd_\infty^1\Sigma'$.
\end{enumerate}
\end{Theorem}

\begin{Theorem}[Richards-Stoilow] The following statements hold:
\begin{enumerate}
\item The unoriented homeomorphism type of an open Riemann surface
$\Sigma$ is entirely determined by the triplet $(g(\Sigma),
\pd_\infty^1\Sigma,\pd_\infty\Sigma)$.
\item Consider any triplet $(g,F^1,F)$ with $g\in \Z_{\geq 0}\sqcup
\{\aleph_0\}$, $F$ a non-empty compact, Hausdorff and totally
disconnected countable topological space and $F^1\subset F$ a
(possibly empty) subset of $F$, endowed with the topology induced from
$F$. Then there exists an open Riemann surface $\Sigma$ with
$g(\Sigma)=g$, $\pd_\infty\Sigma\simeq_{\mathrm{homeo}}F$ and
$\pd_\infty^1\Sigma\simeq_{\mathrm{homeo}}F^1$.
\end{enumerate}
\end{Theorem}

\noindent The following results are crucial in the context of B-type LG models: 

\begin{Theorem}[Behnke-Stein]
Every open Riemann surface is Stein.
\end{Theorem}

\begin{Theorem}[Grauert-R\"ohrl]
Any holomorphic vector bundle on an open Riemann surface $\Sigma$ is
holomorphically trivial. In particular, the analytic Picard group
$\Pic(\Sigma)$ vanishes and the canonical line bundle
$K_\Sigma$ is holomorphically trivial.
\end{Theorem}

\noindent It follows that any open Riemann surface is Stein (and hence
K\"ahlerian) and holomorphically Calabi-Yau. Hence any open Riemann
surface can be used as the target space of a B-type LG model whose TFT
datum admits the simplified description discussed in \cite{lg2} and
summarized in \cite{mirela}. Since the holomorphic tangent bundle $T\Sigma$ is
trivial by Grauert-R\"ohrl theorem, any open Riemann surface is
holomorphically parallelizable.

\begin{Proposition}[various authors]
Every non-compact Riemann surface $\Sigma$ admits a holomorphic
embedding in $\C^3$ and a holomorphic immersion in $\C^2$.
\end{Proposition}

\paragraph{\bf Remark.}
It is not known if any open Riemann surface admits a proper
holomorphic embedding in $\C^2$.

\subsection{The ring $\O(\Sigma)$}

\

\noindent Let $\Sigma$ be an open Riemann surface. This subsection recalls
certain results regarding the ring $\O(\Sigma)$ of complex-valued holomorphic functions
defined on $\Sigma$. 

\begin{Theorem}[Bers]
Let $\Sigma_1$ and $\Sigma_2$ be two connected non-compact Riemann
surfaces. Then $\Sigma_1$ and $\Sigma_2$ are biholomorphic iff their
rings of holomorphic functions $\O(\Sigma_1)$ and $\O(\Sigma_2)$ are
isomorphic {\bf as unital $\C$-algebras}.
\end{Theorem}

\begin{Theorem}[Iss'sa]
Let $\Sigma_1$ and $\Sigma_2$ be two connected non-compact Riemann
surfaces. Then $\Sigma_1$ and $\Sigma_2$ are biholomorphic iff their
fields of meromorphic functions $\cM(\Sigma_1)$ and $\cM(\Sigma_2)$
are isomorphic {\bf as unital $\C$-algebras}.
\end{Theorem}

\begin{Proposition}[Henriksen-Alling]
The following statements hold:
\begin{enumerate}[1.]
\item The cardinal Krull dimensions of all open Riemann surfaces are
  equal to each other (denote this cardinal number by $\bk$).
\item We have $\bk\geq 2^{\aleph_1}$.
\end{enumerate}
\end{Proposition}

\begin{Definition}
An {\em elementary divisor domain} (EDD) is an integral domain $R$
such that {\bf any} matrix with coefficients from $R$ admits a Smith
normal form.
\end{Definition}

\begin{Theorem}[Helmer-Henriksen-Alling]
For any open Riemann surface $\Sigma$, the ring $\O(\Sigma)$ is an
elementary divisor domain.
\end{Theorem}

\subsection{Special uniformizers}

\

\noindent Let $\Sigma$ be an open Riemann surface and
$\ord_x:\cM(\Sigma)\rightarrow \Z$ be the map which assigns to a
meromorphic function $f\in \cM(\Sigma)$ its order at the point $x\in
\Sigma$.

\begin{Definition} 
A {\em special local uniformizer} for $\Sigma$ at the point $x\in
\Sigma$ is a meromorphic function $t_x\in \cM(\Sigma)$ such that
$\ord_x(t_x)=1$ and $\ord_y(t_x)=0$ for all $y\in \Sigma\setminus
\{x\}$. Notice that $t_x\in \O(\Sigma)$. 
\end{Definition}

\begin{Proposition}
The special local uniformizers of $\Sigma$ coincide with the prime
elements of $\O(\Sigma)$. In particular, a special local uniformizer
at $x$ is determined up to multiplication by a unit of $\O(\Sigma)$
{\rm (}i.e. a nowhere-vanishing element of $\O(\Sigma)${\rm )}.
\end{Proposition}

\subsection{Critically-finite elements of $\O(\Sigma)$}

\

\noindent Most of our results for B-type Landau-Ginzburg models with
open Riemann surface target depend on the assumption that the
Landau-Ginzburg superpotential is {\em critically finite} in the sense
discussed below. The notion of critically finite element was
introduced in \cite{bezout,edd}, which studied categories of finite
rank matrix factorizations for B\'ezout and elementary divisor
domains.

\begin{Definition}
A divisor $r\in \O(\Sigma)$ of $f\in \O(\Sigma)$ is called {\em
  critical} if $r^2|f$.
\end{Definition}

\begin{Definition}
A non-zero non-unit $f\in \O(\Sigma)$ is called:
\begin{itemize}
\itemsep 0.0em
\item {\em non-critical}, if $f$ has no critical divisors;
\item {\em critically-finite}, if $f$ has a factorization of the form:
\ben
\label{Wcritform}
f=f_0 f_c~~\mathrm{with}~~f_c=p_1^{n_1}\ldots p_N^{n_N}~~,
\een
where $N\geq 1$, $n_i\geq 2$, $p_1,\ldots, p_N\in \O(\Sigma)$ are
critical prime divisors of $f$ with $(p_i)\neq (p_j)$ for $i\neq j$
and $f_0\in \O(\Sigma)$ is non-critical and coprime with $f_c$.
\end{itemize}
\end{Definition}

\begin{Lemma}
\label{lemma:div} 
Let $f\in \O(\Sigma)$ be a non-zero element and $D(f)$ be the analytic
divisor of $f$. Then the following statements hold:
\begin{enumerate}[1.]  
\itemsep 0.0em
\item $f$ is a unit of $\O(\Sigma)$ iff $D(f)=0$.
\item $f$ is a prime element of $\O(\Sigma)$ (i.e. a special
  local uniformizer of $\Sigma$) iff $D(f)=x$ for some $x\in \Sigma$.
\item $f$ is non-critical iff $D(f)$ is multiplicity-free at any point
  in its support, i.e. iff it has the form: 
\be 
D(f)=\sum_{x\in\cZ(f)}{x}~~,
\ee
where $\cZ(f)$ denotes the zero set of $f$ (which is at most countable).
\item If $g\in \O(\Sigma)$ is another holomorphic function, then $f|g$
  iff $D(f)\leq D(g)$.
\item $f$ is critically-finite iff $f=g h$, where $g\in \O(\Sigma)$ is
non-critical and $h\in \O(\Sigma)$ satisfies:
\be 
D(h)=\sum_{x\in \cZ(h)}n_x \cdot {x}~~, 
\ee 
with $\cZ(h)$ a non-empty finite set and $n_x>1$ for every $x \in \cZ(h)$.
\end{enumerate}
\end{Lemma}

\section{B-type Landau-Ginzburg models with one-dimensional target}

\

\begin{Definition} A {\em Landau-Ginzburg (LG) pair} is a pair $(X,W)$ such
that:
\begin{enumerate}[1.]
\item $X$ is a non-compact complex and K\"ahlerian connected manifold
of dimension $d>0$, which is holomorphically Calabi-Yau in the sense
that its holomorphic canonical line bundle $K_X$ is holomorphically
trivial.
\item $W\in \O(X)$ is a non-constant holomorphic complex-valued
function defined on $X$, which is called the {\em Landau-Ginzburg
superpotential}.
\end{enumerate}
\end{Definition}

\noindent Given an LG pair $(X,W)$, let $Z_W\eqdef \{x\in X| (\pd
W)(x)=0\}$ be the critical set of $W$. Since any Stein manifold is
K\"ahlerian, a particular case of LG pair is obtained by taking $X$ to
be Stein.  This situation was studied in \cite{lg2} (see \cite{mirela}
for a summary).

\begin{Definition} A {\em one-dimensional LG pair} is an LG pair
$(X,W)$ such that $\dim_\C X=1$.
\end{Definition}

\noindent Since any open Riemann surface is Stein and holomorphically
Calabi-Yau (see Subsection \ref{subsec:RScx}), we have:

\begin{Proposition} An LG pair is one-dimensional iff $X$ coincides
with an open Riemann surface~$\Sigma$.
\end{Proposition}

\noindent Since any compact analytic subset of a Stein manifold is
finite, the critical set of $W\in \O(\Sigma)$ is compact iff it is
finite. Let $(\Sigma,W)$ be a one-dimensional LG pair with compact
(and hence finite) critical set $Z_W$. For any $p\in Z_W$, let:
\be
\rM(\hW_p)\eqdef \frac{\cO_{\Sigma,p}}{\langle {\hW}'_p\rangle}
\ee
denote the {\bf analytic} Milnor algebra of the analytic function germ
${\hW}_p$ of $W$ at $p$ and $\nu_p\eqdef \dim_\C\rM({\hat W}_p)$
denote the analytic Milnor number of $W$ at $p$. Here $\langle
W'_p\rangle$ is the ideal of $\O(\Sigma)$ generated by $\frac{\dd
W(p)}{\dd z}$, where $z$ is any local holomorphic coordinate of
$\Sigma$ centered at $p$ (this ideal is independent of the choice of
such a local coordinate). Let $t_p\in \cM(\Sigma)$ be a {\em special} local
uniformizer of $\Sigma$ at $p$.

\begin{Proposition}{\rm \cite{surface}}
We have $\rM({\hat W}_p)\simeq_{\O(\Sigma)} \oplus_{i=1}^N \cO_{\Sigma,p}/\langle {\hat t}_p^{\nu_p}\rangle$~~.
\end{Proposition}

\begin{Proposition}{\rm \cite{surface}}
The bulk state space of the open-closed TFT defined by $(\Sigma,W)$
can be identified with the direct sum of analytic Milnor algebras:
\be
\cH\simeq_{\O(X)} \oplus_{p\in Z_W} \rM({\hat W}_p)~~.
\ee 
In particular, we have:
\be
\dim_\C \cH=\sum_{p\in Z_W}{\nu_p}~~.
\ee
\end{Proposition}

\subsection{The topological D-brane category}

\

\noindent For any unital commutative ring $R$, let $\MF(R,W)$ denote
the category of finite rank matrix factorizations of $W$ over $R$ and
$\HMF(R,W)$ denote its total cohomology category (which is
$\Z_2$-graded and $R$-linear). Let $\hmf(R,W)\eqdef \HMF^\0(R,W)$
denote the sub-category obtained from $\HMF(R,W)$ by keeping only the
even morphisms. It is well-known that $\hmf(R,W)$ has a natural
structure of triangulated category with involutive shift functor.

\begin{Proposition}{\rm \cite{surface}} Let $(\Sigma,W)$ be a
one-dimensional LG pair. Then the D-brane category $\cT$ of the
associated 2d TFT (i.e. of the quantum LG model with target $(X,W)$)
can be identified with:
\be
\cT\simeq_{\O(\Sigma)} \HMF(\O(\Sigma),W)~~.
\ee
\end{Proposition}

\begin{Definition}
A matrix factorization $(M,D)$ of $W$ over
$\O(\Sigma)$ is called {\em elementary} if $\rk M^\0=\rk M^\1=1$.
\end{Definition}

\noindent In the limit $W\rightarrow 0$, the B-type LG model
parameterized by $(X,W)$ reduces to a B-type non-linear sigma model
with non-compact target $X$. Intuitively, elementary factorizations
correspond to `elementary D-branes', i.e. those topological D-branes
which are obtained by 'topological tachyon condensation' in a pair
formed of a single brane and a single anti-brane of this non-linear
sigma model, where tachyon condensation is driven by turning on the
Landau-Ginzburg superpotential $W$. See \cite{sft1,sft2,sft3,HLL} for
the origin of this interpretation in open topological string field
theory.

\begin{Definition} The $\Z_2$-graded {\em cocycle category}
$\ZMF(\O(\Sigma),W)$ has the same objects as $\MF(\Sigma,W)$ but its
morphisms are the {\em closed} morphisms of $\MF(\O(\Sigma),W)$.
\end{Definition}

\noindent Let $\zmf(\O(\Sigma),W)$ be the even subcategory of
$\ZMF(\O(\Sigma),W)$. We say that two matrix factorizations of $W$ are
\emph{strongly isomorphic} if they are isomorphic in
$\zmf(\O(\Sigma),W)$. Notice that an elementary factorization of $W$
is indecomposable in the additive category $\zmf(\O(\Sigma),W)$, but it 
need not be indecomposable in $\cT^\0=\hmf(\O(\Sigma),W)$.

\subsection{Critically-finite superpotentials}

\

\noindent The following result characterizes critically-finite superpotentials:

\begin{Proposition} {\rm \cite{surface}}
Let $(\Sigma,W)$ be a one-dimensional LG pair. Then the
following statements are equivalent:
\begin{enumerate}[(a)] \itemsep 0.0em
\item $W$ is a critically-finite element of $\O(\Sigma)$.
\item The intersection $\cZ(W)\cap Z_W$ is finite.
\item The divisor of $W$ has the form: 
\be 
D(W)=D_0+\sum_{i=1}^N{n_i  x_i}~~, 
\ee 
where $D_0$ is either the trivial divisor or an effective divisor
whose multiplicity at every point of its support is one, the symbols
$x_1,\ldots, x_N$ (with $N\geq 1$) denote a finite collection of
points of $\Sigma$ which do not belong to the support of $D_0$ and
$n_i\geq 2$ for all $i=1,\ldots, N$.
\item We have $W=W_0W_c$, where $W_0$ has no zeros or only simple
zeros, $W_c$ has a finite number of zeros, each of which has
multiplicity at least two and $W_0$, $W_c$ have no common zero.
\end{enumerate}
In particular, any holomorphic function $W$ with finite critical set
is a critically-finite element of $\O(\Sigma)$.
\end{Proposition}

\paragraph{\bf Remark.}
Any critically-finite superpotential $W\in
\O(\Sigma)$ can be written as:
\begin{equation}
\label{Wcf}
W=W_0W_c~~,~\mathrm{with}~~W_c=\prod_{i=1}^N t_{x_i}^{n_i}~~,
\end{equation}
where: 
\begin{enumerate} 
\itemsep 0.0em
\item $W_0\in \O(\Sigma)$ has a finite or countable number of zeros,
all of which are simple.
\item $x_1,\ldots, x_N$ (with $N\geq 1$) are distinct points of
$\Sigma$ and $t_{x_i}$ are special uniformizers at these points.
\item None of the points $x_i$ is a zero of $W_0$.
\end{enumerate} In this case, we have
$D_0=D(W_0)$ and we set $D_c\eqdef \sum_{i=1}^N{n_i x_i}$.

\subsection{Primary factorizations and the Krull-Schmidt property of $\hmf(\O(\Sigma),W)$}

\

\noindent Recall that an additive category $\cA$ is called {\em Krull-Schmidt}
if any object of $\cA$ decomposes as a finite sum of indecomposable
objects with quasi-local endomorphism rings. We say that a matrix
factorization of $W$ is {\em trivial} if it is a zero object in
$\hmf(\Sigma,W)$.

\begin{Proposition}{\rm \cite{surface}}
 Any elementary factorization of $W$ over $\O(\Sigma)$ is strongly
 isomorphic to an elementary factorization of the form $e_v\eqdef
 (\O(\Sigma)^{1|1},D_v)$, where $v\in \O(\Sigma)$ is a divisor of $W$
 and $D_v\eqdef \left[\begin{array}{cc} 0 & v\\ u &
     0 \end{array}\right]$, with $u\eqdef W/v\in \O(\Sigma)$.
\end{Proposition}

\begin{Definition}
An element $f\in \O(\Sigma)$ is called {\em primary} if it is a power
of a prime element of $\O(\Sigma)$. An elementary factorization $e_v$
of $W$ is called {\em primary} if $v$ is a primary divisor of $W$.
\end{Definition}

\noindent If $e_v$ is a primary factorization of $W$ over
$\O(\Sigma)$, then we have $v=a t_x^k$ for some $a\in
\O(\Sigma)^\times$, some $k\in \Z_{>0}$ and some $x\in \Sigma$. In
this case, $W$ has a zero of order at least $k$ at the point $x\in
\Sigma$.

\begin{Theorem}{\rm \cite{surface}}
Suppose that $W$ is critically-finite. Then the additive category
$\hmf(\O(\Sigma),W)$ is a Krull-Schmidt category whose non-zero
indecomposables are the nontrivial primary matrix factorizations of
$W$.
\end{Theorem}

\noindent Hence any finite rank matrix factorization (in particular,
any non-primary elementary factorization) of $W$ decomposes into
primary factorizations in the category $\hmf(\O(\Sigma),W)$. The
primary factorizations of $W$ play the role of `simple' (a.k.a `irreducible')
topological D-branes of the B-type LG model.

\subsection{The triangulated structure of $\hmf(\O(\Sigma),W)$}

\

\noindent Let $R$ be a unital commutative ring. Let $\mod_R$ be the Abelian category of
finitely-generated $R$-modules. 

\

\begin{Definition}

\

\begin{enumerate}
\item The {\em projectively stable category} $\umod_R$ has the same
objects as $\mod_R$ and modules of morphisms given by:
\be
\uHom_R(M,N) \eqdef \Hom_R(M,N)/\mathcal{P}_R(M,N)~~~ \forall M,N \in \Ob(\mod_R)~,
\ee
where $\mathcal{P}_R(M,N) \subset \Hom_R(M,N)$ consists of those
morphisms of $\mod_R$ which factor through a projective module of
finite rank. 
\item The {\em injectively stable category}
$\overline{\mod}_R$ has the same objects as $\mod_R$ and modules of
morphisms given by:
\be
\overline{\Hom}_R(M,N) \eqdef \Hom_R(M,N)/\mathcal{I}_R(M,N)~~~ \forall M,N \in \Ob(\mod_R)~,
\ee
where $\mathcal{I}_R(M,N) \subset \Hom_R(M,N)$ consists of those
morphisms of $\mod_R$ which factor through an injective module of
finite rank. 
\end{enumerate}
\end{Definition}

\paragraph{\bf Remark.}  When $R$ is a self-injective ring, the
categories $\umod_R$ and $\overline{\mod}_R$ are equivalent to each
other and the category $\umod_R$ is naturally triangulated since in
this case $\mod_R$ is a Frobenius category (i.e. an exact category
whose projective and injective objects coincide).

\

\begin{figure}[H] \centering \scalebox{0.7}{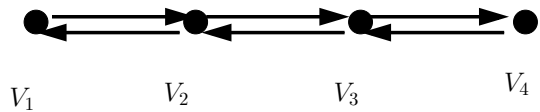}
\caption{Auslander-Reiten quiver for $\umod_{A_i}$ when $n_i=5$. The
  Auslander-Reiten translation fixes all vertices and the
  multiplicities of all arrows are trivial.}
  \label{squiver}
\end{figure}

\noindent The following result gives a complete description of the
triangulated structure of the category $\cT^\0=\hmf(\O(\Sigma),W)$ for the
case when $W$ is critically finite:
  
\begin{Theorem}{\rm \cite{surface}}
\label{thm:cT}
Let $W\in \O(\Sigma)$ be a critically-finite superpotential of the
form \eqref{Wcf}. Then the following statements hold:
\begin{enumerate}[1.]  
\itemsep 0.0em
\item The ring $A_i\eqdef \O(\Sigma)/\langle t_{x_i}^{n_i}\rangle$ is
  Artinian and Frobenius (hence also self-injective) for all
  $i=1,\ldots, N$.
\item For each $i=1,\ldots, N$, there exist equivalences of
  triangulated categories:
\be
\rD_\Sing(A_i)\simeq \umod_{A_i}~~,
\ee
where $\umod_{A_i}$ denotes the projectively stable category of
finitely-generated $A_i$-modules.
\item The triangulated category $\umod_{A_i}$ is Krull-Schmidt with
  non-zero indecomposable objects given by the $A_i$-modules:
\be
V_i^{(k)}\eqdef \O(\Sigma)/\langle t_{x_i}^k\rangle\simeq \langle t_{x_i}^{n_i-k}\rangle/
\langle t_{x_i}^{n_i} \rangle~~\mathrm{where}~~k=1,\ldots, n_i-1~~.
\ee
This category admits Auslander-Reiten triangles, having the
Auslander-Reiten quiver shown in Figure \ref{squiver}. Moreover,
it is classically generated by the residue field
$V_i^{(1)}=\O(\Sigma)/\langle t_{x_i}\rangle$ of $A_i$.
\item The triangulated category $\rD_\Sing(A_i)\simeq \umod_{A_i}$ is
$1$-Calabi-Yau for all $i$, with involutive shift functor:
\be
\rOmega(V_i^{(k)})=V_i^{(n_i-k)}~~~\forall k=1,\ldots, n_i-1~~.
\ee
\item There exists an equivalences of triangulated categories:
\ben
\label{tequiv} 
\hmf(\O(\Sigma),W) \simeq \vee_{i=1}^N \rD_\Sing(A_i)\simeq \vee_{i=1}^N \umod_{A_i}~~.
\een
\item The triangulated category $\hmf(\O(\Sigma),W)$ is $1$-Calabi-Yau,
Krull-Schmidt and admits Auslander-Reiten triangles. Its
Auslander-Reiten quiver is disconnected, with connected components
given by the Auslander-Reiten quivers of the categories
$\umod_{A_i}$.
\end{enumerate}
\end{Theorem}

\

\subsection{Counting indecomposables in $\hmf(\O(\Sigma),W)$}

\

\noindent The following result gives the count of simple topological
D-branes of the LG model:

\begin{Proposition}{\rm \cite{surface}}
Suppose that $W$ is critically finite of the form given in
eq. (1). Then the number of isomorphism classes of indecomposable
non-zero objects of the category $\hmf(\O(\Sigma),W)$ equals
$\sum_{i=1}^N (n_i-1)=-N+\sum_{i=1}^N n_i$.
\end{Proposition}

\noindent The degrees $n_i$ of the
prime factors $t_{x_i}$ arising in the decomposition of
$W_c$ define a $\Z_2$-grading on the set $I_N\eqdef \{1,\dots,N\}$
whose components are given by:
\ben
I_N^\0\eqdef \{i\in I_N \,\big| \, n_i~\mathrm{is~even}\}~,~I_N^\1\eqdef \{i\in I_N\,\big|\, n_i~\mathrm{is~odd}\}~.
\een
Let:
\be
N^{\0}\eqdef |I_N^{\0}|~~\mathrm{and}~~N^{\1}\eqdef |I_N^{\1}|~~
\ee
denote the cardinalities of these subsets of $I$, which satisfy
$N^\0+N^\1=N$. Any non-empty subset $K\subset I_N$ is endowed with the
$\Z_2$-grading induced from $I_N$, which has components:
\be
K^\0\eqdef K\cap I_N^\0~~,~~K^\1\eqdef K\cap I_N^\1~~.
\ee

\subsection{Counting elementary factorizations in $\hmf(\O(\Sigma),W)$ and $\HMF(\O(\Sigma),W)$}

\

\noindent The following result allows one to count elementary
topological D-branes as well as elementary brane-antibrane pairs:

\begin{Theorem}{\rm \cite{surface}}
Suppose that $W$ is a critically-finite element of $\O(\Sigma)$ with
the decomposition given in eq. (1). Then: 
\begin{enumerate}[1.]
\itemsep 0.0em
\item The number of isomorphism classes of elementary factorizations
in the category $\hmf(\O(\Sigma),W)$ is given by:
\ben
\label{Nfinalc}
{\widehat \cN}_W=\sum_{k=0}^{N^\1}\sum_{\substack{K\subsetneq I_N, \\|K^\1|=k}} 2^{N^{\0}+k} \prod_{i\in K} \big\lfloor\frac{n_i-1}{2}\big\rfloor ~~.
\een
\item The number of isomorphism classes of elementary matrix
factorizations in the category $\HMF(\O(\Sigma),W)$ is given by:
\ben
\label{Nfinal}
\cN_W=2^{r^\0}+ \sum_{k=0}^{N^\1} 2^{N^{\0}+k-1} \sum_{\substack{K\subsetneq I_N \\|K^\1|=k}}\prod_{i\in K} \big\lfloor\frac{n_i-1}{2}\big\rfloor ~~.
\een
\end{enumerate}
\end{Theorem}

\ack{

\

\noindent This work was supported by the research grants IBS-R003-S1
(``Constructive string field theory of open-closed topological B-type
strings'') and IBS-R003-D1. The second author was also supported by
the Max Planck Institute for Mathematics, Bonn and by the Australian
Research Council grant DP180103891.}

\section*{References}

\


\begin{thebibliography}{10}
\bibitem{LG1}{Lazaroiu, C. I., On the boundary coupling of topological
Landau-Ginzburg models, {\it JHEP} {\bf 0505} (2005) 037.}
\bibitem{LG2}{Herbst, M., Lazaroiu, C. I., Localization and traces in
open-closed topological Landau-Ginzburg models, {\it JHEP} {\bf 0505}
(2005) 044.}
\bibitem{top}{Lazaroiu, C. I., On the structure of open-closed
topological field theory in two dimensions, {\it Nucl. Phys. B} {\bf
603} (2001) 497--530.}
\bibitem{lg1}{Babalic, E. M., Doryn, D., Lazaroiu, C. I., Tavakol, M.,
Differential models for B-type open-closed topological Landau-Ginzburg
theories, {\it Commun. Math. Phys.} {\bf 361} (2018) 1169--1234 (arXiv:1610.09103v3 [math.DG]).}
\bibitem{tserre}{Doryn, D., Lazaroiu, C. I., Non-degeneracy of
cohomological traces for general Landau-Ginzburg models, {\it
preprint}, arXiv:1802.06261 [math.AG].}
\bibitem{lgproc}{Babalic, E. M., Doryn, D., Lazaroiu, C. I., Tavakol, M., A
differential model for B-type Landau-Ginzburg theories, {\it
preprint}, arXiv:1709.00684 [math.DG].}
\bibitem{lg2}{Babalic, E. M., Doryn, D., Lazaroiu, C. I., Tavakol, M., On
B-type open-closed Landau-Ginzburg theories defined on Calabi-Yau
Stein manifolds, {\it Commun. Math. Phys.} {\bf 362} (2018) 129--165 (arXiv:1610.09813v3 [math.DG]).}
\bibitem{mirela}{Babalic, E. M., Doryn, D., Lazaroiu, C. I., Tavakol, M.,
B-type Landau-Ginzburg models on Stein manifolds, {\it preprint},
conference proceedings for ``Group 32'', Prague, July 9-13, 2018.}
\bibitem{surface}{Lazaroiu, C. I., Tavakol, M., B-type topological
Landau-Ginzburg models over general non-compact Riemann surfaces, {\it
preprint available at https://cgp.ibs.re.kr/archive/preprints} .}
\bibitem{bezout}{Doryn, D., Lazaroiu, C. I., Tavakol, M., Elementary
matrix factorizations over B\'ezout domains, {\it preprint},
arXiv:1801.02369 [math.AC].}
\bibitem{edd}{Doryn, D., Lazaroiu, C. I., Tavakol, M., Matrix
factorizations over elementary divisor domains, {\it preprint},
arXiv:1802.07635 [math.AC].}
\bibitem{Kerekjarto}{Ker\'ekj\'art\'o, B., Vorlesungen \"uber Topologie
I, Springer, Berlin, 1923.}
\bibitem{Stoilow}{Stoilow, S., Le\c{c}ons sur les principes
topologiques de la th\'eorie de fonctions analytiques, 2nd ed.,
Gauthier- Villars, Paris, 1956.}
\bibitem{Richards}{Richards, I., On the Classification of Non-Compact
Surfaces, {\it Transactions of the AMS} {\bf 106} (1963) 2, 259--269.}
\bibitem{sft1}{Lazaroiu, C. I., Generalized complexes and string field
theory, {\em JHEP} {\bf 0106} (2001) 052.}
\bibitem{sft2}{Lazaroiu, C. I., Unitarity, D-brane dynamics and D-brane
categories, {\em JHEP} {\bf 0112} (2001) 031.}
\bibitem{sft3}{Lazaroiu, C. I., D-brane categories, {\em
Int. J. Mod. Phys.} {\bf A18} (2003) 5299--5335.}
\bibitem{HLL}{Herbst, M., Lazaroiu, C. I., Lerche, W., D-brane effective
action and tachyon condensation in topological minimal models, {\em
JHEP} {\bf 0503} (2005) 078.}
\end{thebibliography}
\end{document}